\documentclass[12pt,preprint]{aastex}

\def  \be   {\begin{equation}}
\def  \ee   {\end{equation}}
\def  \beq  {\begin{eqnarray}}
\def  \eeq  {\end{eqnarray}}

\shorttitle{Magnetic dynamo action in astrophysical turbulence}
\shortauthors{Malyshkin \& Boldyrev}

\begin{document}

\title{Magnetic dynamo action in astrophysical turbulence}

\author{Leonid Malyshkin\altaffilmark{1} and Stanislav Boldyrev\altaffilmark{2}}
\affil{${~}^1$Department of Astronomy \& Astrophysics,
University of Chicago, 5640 S. Ellis Ave., Chicago, IL 60637; {\sf leonmal@uchicago.edu}}
\affil{${~}^2$Department of Physics, University of Wisconsin-Madison, 1150 University Ave.,
Madison, WI 53706; {\sf boldyrev@wisc.edu}}

%------------------------------------------------------------------------------------------

\begin{abstract}
We investigate the structure of magnetic field amplified by turbulent
velocity fluctuations, in the framework of the kinematic
Kazantsev-Kraichnan model. We consider Kolmogorov distribution of
velocity fluctuations, and assume that both Reynolds number and magnetic
Reynolds number are very large. We present the full numerical solution of the model for 
the spectra and the growth rates of magnetic fluctuations. We consider astrophysically 
relevant limits of large and small magnetic Prandtl numbers, and address 
both helical and nonhelical cases.
\end{abstract}

%------------------------------------------------------------------------------------------

\section{Introduction}

Magnetic fields are found everywhere in the universe, in planets and stars,
in galaxies and galaxy clusters. Magnetic fields in astrophysical systems are usually generated 
in a broad interval of scales, ranging from small resistive scales  
to large scales exceeding the correlation length of plasma motions.  One of the most important, 
challenging and still open
questions in astrophysics is how cosmic magnetic fields have been generated
and what is their structure. The prevailing theory for the origin of magnetic
fields is dynamo action, which is stretching of magnetic field lines
by random motion of highly conducting plasmas or fluids in which these lines are frozen
\citep[e.g.,][]{VZ_72,parker,lynden-bell,zweibel,brandenburg2,kulsrud1,schekochihin,kulsrud2}. 
Small-scale magnetic fields, that is fields at the scales smaller than the velocity field scales,   
are generally expected under this mechanism, while  large-scale magnetic fields, correlated 
at scales larger than the correlation
scale of a velocity field, can be generated if some additional conditions are met, 
such as the condition that the velocity field ${\bf v}({\bf x, t})$ is not mirror symmetric,
say, possesses nonzero kinetic 
helicity $H=\int {\bf v}\cdot ({\nabla \times {\bf v}})\, d^3 x\neq 0$ \citep{skr,moffatt78}. 

Assume that the velocity fluctuations have the typical correlation scale~$l_0$, 
and the typical rms value~$v_0$, and the plasma has kinematic viscosity~$\nu$ and 
magnetic diffusivity~$\eta$ (the latter is proportional to electrical resistivity). 
The range of scales available for velocity and magnetic 
fluctuations can be characterized by the Reynolds number ${\rm Re}\sim l_0v_0/\nu$ and the magnetic 
Reynolds number ${\rm Rm}\sim l_0v_0/\eta$, respectively. In astrophysical applications both numbers 
are very large (for example, in a protogalaxy, where dynamo action is believed to operate, 
${\rm Re}$ and ${\rm Rm}$ reach values $\sim 10^{5}$ and $\sim 10^{26}$, respectively). 
Their ratio, the magnetic Prandtl number ${\rm Pm=Rm/Re}$ can be either large or small. For 
example, in galaxies 
and galaxy clusters ${\rm Pm}\gg 1$, while in planets and stellar interiors
${\rm Pm}\ll 1$. Because of the vast range of scales available for magnetic
and velocity fluctuations, and generally strongly disparate magnetic
and velocity dissipation scales, present-day direct numerical simulations cannot 
directly address astrophysical magnetohydrodynamic (MHD) regimes. Indeed, maximal
Reynolds and magnetic Reynolds numbers accessible with numerical simulations are
hopelessly small, of the order of $10^3-10^4$, in which case the typical magnetic 
Prandtl numbers are not significantly different from ${\rm Pm}\sim 1$. A
physical picture of magnetic dynamo action and an effective analytic
framework for investigating astrophysical dynamo action are therefore
in demand.   

The first step in understanding dynamo action is to understand the
initial, kinematic stage of magnetic field amplification. In this
regime, the magnetic field is weak and does not affect the velocity
fluctuations. Magnetic field evolution is therefore fully described by
the induction equation in which the velocity field is prescribed
independently of the magnetic field. An effective framework in this
case is provided by the so-called Kazantsev-Kraichnan model, where the
velocity field is assumed to be a random Gaussian short-time-correlated 
field. This formal simplification allows for analytic solutions of
the model while capturing the essential physics of the phenomenon. We
should note however that even with this simplification the model is
nontrivial and its general solution is not known.  Only certain
special cases have been solved so far, which reveal a good 
agreement with numerical simulations in the parameter 
range accessible to numerics \citep[e.g.,][]{maron,haugen,boldyrev-cattaneo}. 

The Kazantsev-Kraichnan dynamo model
allows one to answer the fundamental questions concerning the
possibility of turbulent dynamo action for given Reynolds and magnetic
Reynolds numbers, the spectrum of growing magnetic fluctuations, the
conditions for large-scale magnetic field amplification, etc. In many
instances, the results obtained in high-resolution numerical
simulations were predicted by the model well before such simulations
become available. We therefore believe that the model can provide a
valuable insight into astrophysical dynamo regimes that can hardly be
accessed through direct numerical simulations in foreseeable future.     

Our preliminary results aimed at the full numerical characterization
of the kinematic dynamo action in the
Kazantsev-Kraichnan framework  were presented in~\citet{malyshkin}. In
particular, we investigated the growth rates of magnetic fluctuations
in the velocity field with the Kolmogorov energy spectrum.   
In the present paper, we explore the spatial structure of the growing
magnetic eigenmodes. We assume the Kolmogorov spectrum of velocity
fluctuations and address extremely large Reynolds numbers, up to~${\rm
Re}\sim 10^9$. This allows us to study astrophysically relevant limits
of very large and very small magnetic Prandtl numbers. For
completeness, we perform our analysis for the cases of both large and
small kinetic helicity.  
In the helical case we also discuss the relevance of the conventional
$\alpha$-model for the description of large-scale dynamo action, in the
case when small-scale magnetic fluctuations are amplified as well.  

In the next section, we describe the Kazantsev-Kraichnan model of
kinematic dynamo action. In Section~3, we present our results, and in
Section~4 we give our conclusions.

\section{Kazantsev-Kraichnan model}

Kinematic dynamo action is described by the induction equation for the magnetic field:
\begin{eqnarray}
\partial_t {\bf B}=\nabla \times ({\bf v}\times {\bf B})+\eta \nabla^2 {\bf B},
\label{induction}
\end{eqnarray}
where ${\bf v}({\bf x}, t)$ is the velocity field, 
${\bf B}({\bf x},t)$ is the magnetic field, and  $\eta$ is the magnetic
diffusivity. In this equation the velocity field is prescribed 
independently of the magnetic field. Following  
\citet{kazantsev68} and \citet{kraichnan68}, we assume that the
velocity field is statistically homogeneous and isotropic and has a
Gaussian distribution with zero mean, 
$\langle{\bf v}\rangle=0$, and the following covariance tensor 
\beq
\langle {v^i}({\bf x},t){v^j}({\bf x}',t') \rangle \!=\!
\kappa^{ij}({\bf x}-{\bf x}')\delta(t-t'),
\label{V_V_TENSOR}
\eeq
where $\kappa^{ij}$ is an isotropic tensor of turbulent diffusivity,
\beq
\kappa^{ij}({\bf x})\!=\!\kappa_N
\left(\delta^{ij}-\frac{x^ix^j}{x^2}\right)+
\kappa_L \frac{x^ix^j}{x^2}+g\epsilon^{ijk}x^{k}.
\label{KAPPA}
\eeq
Here, functions $\kappa_L(x)$ and $g(x)$ describe kinetic energy and helicity,
$x=|{\bf x}|$, brackets $\langle \rangle$ denote ensemble average,
$\epsilon^{ijk}$ is the unit antisymmetric pseudotensor and
summation over repeated indices is assumed.
The first two terms on the right-hand side of Equation~(\ref{KAPPA})
represent the mirror-symmetric, nonhelical part, while the last term
describes the helical part of the velocity fluctuations.
For an incompressible velocity field (the only case we consider here),
we have $\kappa_N(x)=\kappa_L(x)+x\kappa'_L(x)/2$, where the prime
denotes  derivative with respect to~$x=|{\bf x}|$. Therefore,
to describe the velocity field, we specify only two independent functions,
$\kappa_L(x)$ and $g(x)$. The Fourier transformation of
Equation~(\ref{KAPPA}) is
\beq
\kappa^{ij}({\bf k})=F(k)\left(\delta^{ij}-\frac{k^ik^j}{k^2}\right)
+iG(k)\epsilon^{ijl}k^l.
\label{KAPPA_FOURIER}
\eeq
Functions $F(k)$ and $G(k)$ can be obtained from functions $\kappa_L(x)$
and $g(x)$, and vice versa, by using the three-dimensional Fourier
transforms~\citep{monin71}.

The correlator of homogeneous and isotropic magnetic field can similarly be
expressed as
\begin{eqnarray}
\langle B^i({\bf x}, t)B^j(0,t)\rangle =
M_N\left(\delta^{ij}-\frac{x^ix^j}{x^2}\right)
+ M_L\frac{x^ix^j}{x^2}+K\epsilon^{ijk}x^k,
\label{B_B_TENSOR}
\end{eqnarray}
where the field solenoidality constraint ${\rm div}\,{\bf B}=0$ implies
$M_N(x,t)=M_L(x,t)+(x/2)M'_L(x,t)$. To fully describe the magnetic field
correlator, we need to find only two functions, $M_L(x, t)$
and $K(x, t)$, corresponding to magnetic energy and magnetic helicity.
The Fourier transformed version of Equation~(\ref{B_B_TENSOR}) is
\beq
\langle B^i({\bf k},t)B^{*j}({\bf k},t)\rangle =
F_B(k,t)\left(\delta^{ij}-\frac{k^ik^j}{k^2}\right)
- i\frac{H_B(k,t)}{2k^2}\epsilon^{ijl}k^l,
\label{MAGNETIC_SPECTRA}
\eeq
where $F_B(k,t)$ is the magnetic energy spectral function,
$\langle|{\bf B}({\bf k},t)|^2\rangle=2F_B(k,t)$, and $H_B(k,t)$
is the spectral function of the electric current helicity,
$\langle {B^i}^*({\bf k},t)\: i\epsilon^{ijl}k^jB^l({\bf k},t)\rangle=H_B(k,t)$.
The problem is then to find the correlation function (Equation~(\ref{B_B_TENSOR})) 
of the magnetic field, or, alternatively, its Fourier version 
(Equation~(\ref{MAGNETIC_SPECTRA})).

Suppose that the velocity field (Equations~(\ref{V_V_TENSOR}) and~(\ref{KAPPA})) is 
given, i.e., kinetic energy~$\kappa_L(x)$ and kinetic helicity~$g(x)$ are given.
In this case, to find the properties of the growing magnetic field driven by helical
dynamo action, one needs to solve two coupled partial differential equations for
functions $M_L(x,t)$ and $K(x, t)$ related to magnetic energy and magnetic helicity.
Such equations were first derived by \citet{vainshtein}. Due to their complexity,
there have been few theoretical results obtained for the helical
dynamo~\citep[e.g.,][]{kulsrud,kim,blackman}. For direct numerical 
simulations see \citet[][]{brandenburg2} and references therein. 
Recently, it has been established in \citet{boldyrev05} that \citet{vainshtein}
equations also possess a self-adjoint structure, which is similar to a two-component
quantum mechanical ``spinor'' form with imaginary time. These two coupled self-adjoint
differential equations are linear and homogeneous, and they describe the growth of the
magnetic field in the Kazantsev-Kraichnan model with nonzero kinetic helicity. We
solve them numerically by the fourth-order Runge-Kutta integration method and by
matching the numerical solution to the analytical asymptotic solutions at
$x\to0$ and $x\to\infty$, for details see \citet{malyshkin}.

We are interested in fast exponentially growing eigenmodes of the magnetic field amplified
by helical kinematic dynamo. Therefore, both magnetic correlator functions $M_L(x,t)$
and $K(x, t)$ are assumed to be proportional to $\exp(\lambda t)$, where $\lambda$
is the growth rates of the field eigenmodes. It is important that \citet{boldyrev05}
helical dynamo equations, which we solve, are self-adjoint because this guarantees
that all growth rates $\lambda$ are real.
It turns out that in analogy with quantum mechanics, there are two types of
magnetic field eigenmodes: bound (spatially localized) and unbound
(spatially nonlocalized).
First, for growth rates $\lambda>\lambda_0\equiv g^2(0)/[\kappa_L(0)+2\eta]$ the
eigenfunctions are bound and correspond to ``particles'' trapped by the potential
provided by velocity fluctuations. The bound eigenmodes have discrete growth rates,
i.e., $\lambda=\lambda_n>\lambda_0$ where $n=1,2,3...$. The bound eigenfunctions 
decline exponentially to zero as $x\to\infty$.
Second, for $\lambda\le\lambda_0$ the eigenfunctions are unbound and correspond to
``traveling particles''. The unbound eigenmodes have continuous eigenvalues
of their growth rates, $0<\lambda\le\lambda_0$. The unbound eigenfunctions
asymptotically become a mixture of cosine and sine standing waves as $x\to\infty$.
Eigenvalue $\lambda_0$ corresponds to the fastest growing unbound eigenmode.
The structure and other properties of the magnetic field amplified by dynamo action
are fully determined by all growing eigenmodes of the magnetic field. In particular, the
magnetic energy spectral function $F_B(k,t)$ is the sum over all energy spectral
eigenfunctions,
\beq
\langle|{\bf B}({\bf k},t)|^2\rangle/2=F_B(k,t)=
\sum\limits_{n=1}^{n_{\rm max}}c_nF_{B,n}(k,t)+
\int_0^{\lambda_0}c(\lambda)F_{B,\lambda}(k,t)\,d\lambda.
\label{F_B}
\eeq
In this equation $F_{B,n}(k,t)$ and $F_{B,\lambda}(k,t)$ are the energy spectral
eigenfunctions for the bound and unbound eigenmodes, respectively; coefficients
$c_n$ and $c(\lambda)$ depend on the seed magnetic field at the initial moment
when dynamo started to operate.\footnote{
Note that while the energy spectrum $F_B(k,t)=\langle|{\bf B}({\bf k},t)|^2\rangle/2$
stays always positive, the individual energy spectral eigenfunctions can be negative.
The sharp declines in individual eigenfunctions $F_B(k)$ and $H_B(k)$ plotted in 
Figures~\ref{FIGURE_2} and~\ref{FIGURE_3} are due to the logarithmic representation 
of the eigenfunctions in these figures, which has difficulty to suit with values 
$F_B(k)=0$ and $H_B(k)=0$.}
Similarly to Equation~(\ref{F_B}), the electric current helicity spectral function
$H_B(k,t)$ is the sum over all helicity spectral eigenfunctions with the same coefficients
as in Equation~(\ref{F_B}). To find the correlation function of the magnetic field, 
it is sufficient to find all growth rates and the corresponding spectral eigenfunctions.

\section{Results}

To study a case relevant to astrophysical systems, we consider velocity correlation
tensor (Equation~(\ref{KAPPA_FOURIER})) with the Kolmogorov power velocity spectrum and
large Reynolds number, ${\rm Re}\gg1$.
In the Kolmogorov turbulence, the turbulent diffusivity, given by Equation~(\ref{KAPPA}),
scales as $v_ll\sim l^{4/3}$, where $l=|{\bf x}-{\bf x}'|$ \citep[see, e.g.,][]{frisch}.
As a result, the Kolmogorov scaling implies
$\kappa_L(x)\approx\kappa(0)(x/l_0)^{4/3}\approx v_0 l_0(x/l_0)^{4/3}$.
Without loss of generality we take $l_0\sim 1$, $v_0\sim 1$, and therefore
\beq
\begin{array}{lcl}
F(k)&=&k^{-13/3},\\
G(k)&=&-hk^{-1}F(k),
\end{array}
\quad
2\le k\le k_{\rm max}.
\label{POWER_V}
\eeq
Here, the lower cutoff wavenumber is $k_{\rm min}=2$, the upper cutoff
wavenumber $k_{\rm max}\approx2[\kappa_L(0)/\nu]^{3/4}\approx4\nu^{-3/4}$
is determined by the plasma kinematic viscosity $\nu$, and the helicity
parameter $h$ must satisfy the realizability condition $-1\le h\le1$;
the velocity field is maximally helical when $|h|=1$.\footnote{
Given $\kappa_L(x)$, function $g(x)$ cannot be chosen arbitrarily, its Fourier image
must satisfy the realizability condition $|G(k)|\le F(k)/k$~\citep{moffatt78}.
This results in the condition $-1\le h\le1$. Analogously, given $M_L(x,t)$, function
$K(x,t)$ is restricted by condition $|H_B(k,t)|\le 2kF_B(k,t)$.
}

\begin{figure}[t]
%\epsscale{1.0}
%\plotone{f1.eps}
\vspace{5.8truecm}
\includegraphics{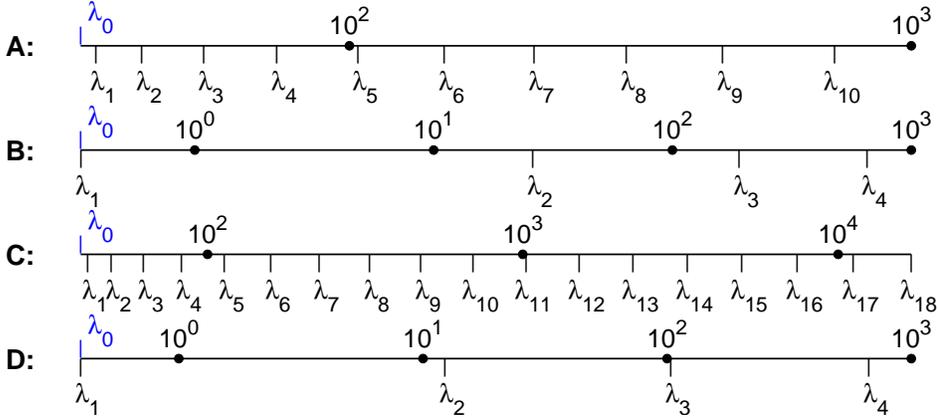}
\caption{Growth rates $\lambda_n$ of the bound magnetic eigenmodes, and $\lambda_0$
of the fastest growing unbound eigenmode. Plots~(A) and~(B) are for $h=1$ and $0.1$,
respectively, while $k_{\rm max}=3000$ (Reynolds number ${\rm Re}\gg1$ and Prandtl
number ${\rm Pm}\gg1$). Plots~(C) and~(D) are for $h=1$ and $0.1$, respectively, while
$k_{\rm max}=3\times10^7$ (${\rm Re}\gg1$ and ${\rm Pm}\ll1$). All plots are
on the logarithmic scale.
\label{FIGURE_1}
}
\end{figure}

\begin{figure}[t]
%\epsscale{1.0}
%\plottwo{f2_UpperLeft_3e3_1.0.eps}{f2_UpperRight_3e3_0.1.eps}
%\plottwo{f2_LowerLeft_3e7_1.0.eps}{f2_LowerRight_3e7_0.1.eps}
\vspace{11.8truecm}
\includegraphics{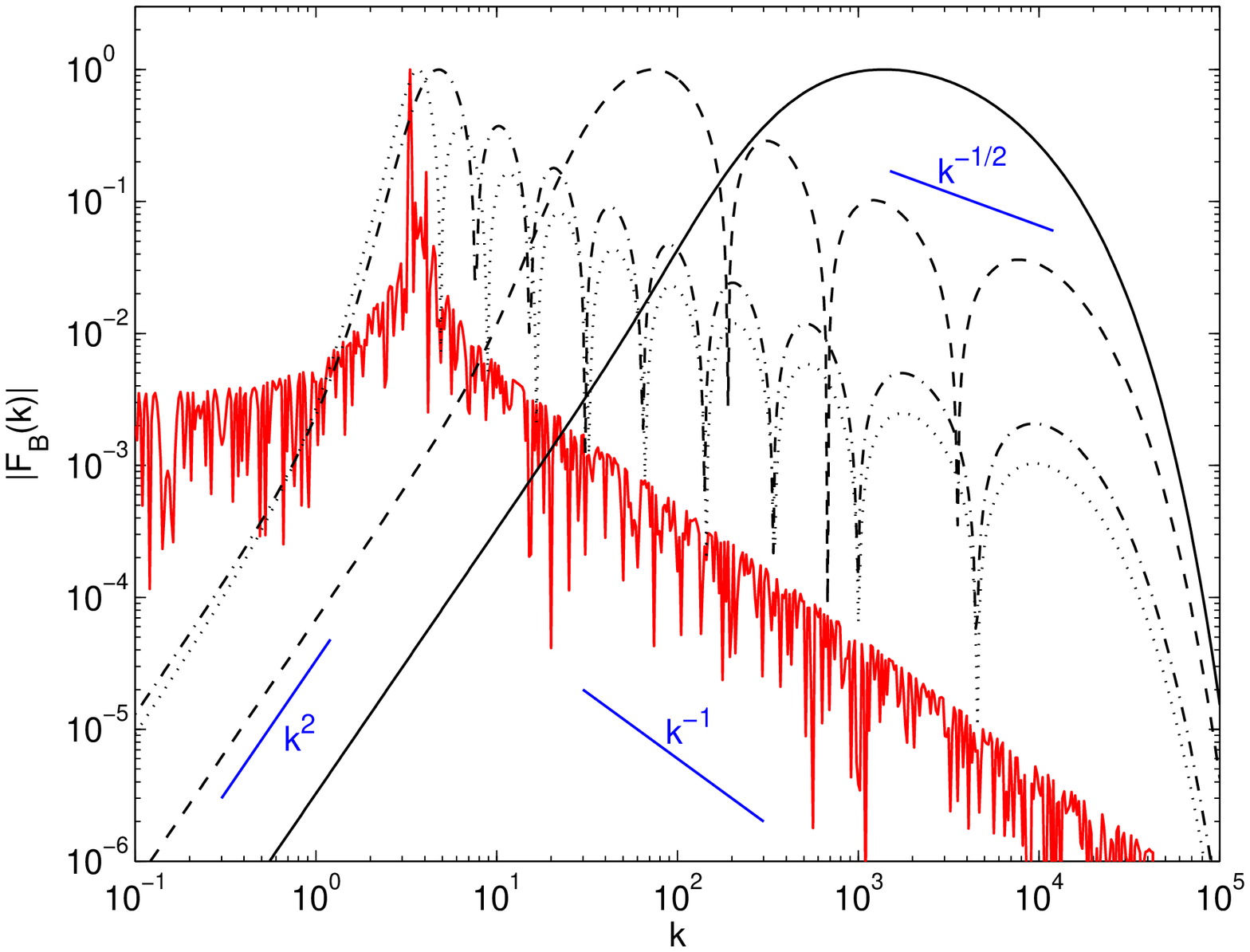}
\includegraphics{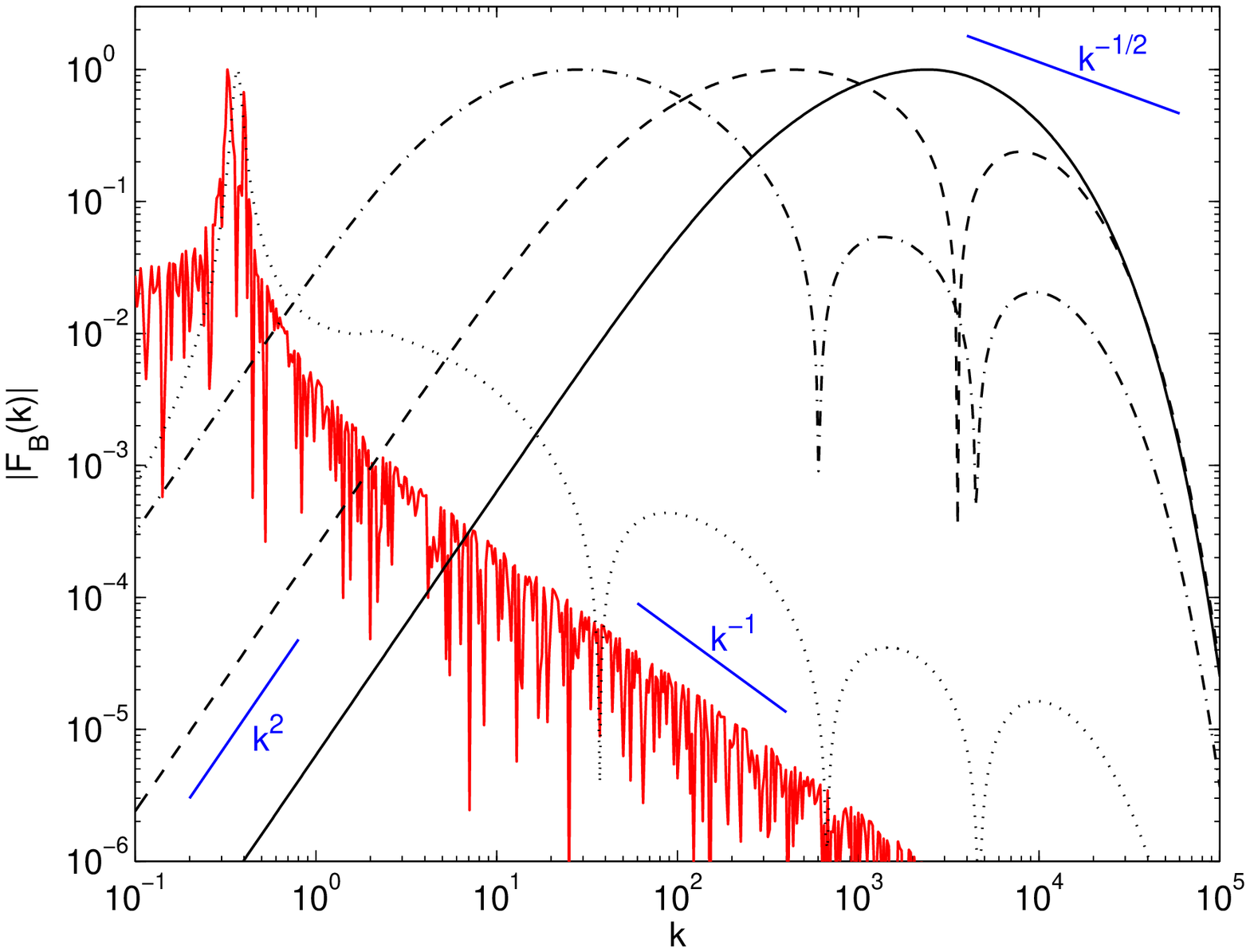}
\includegraphics{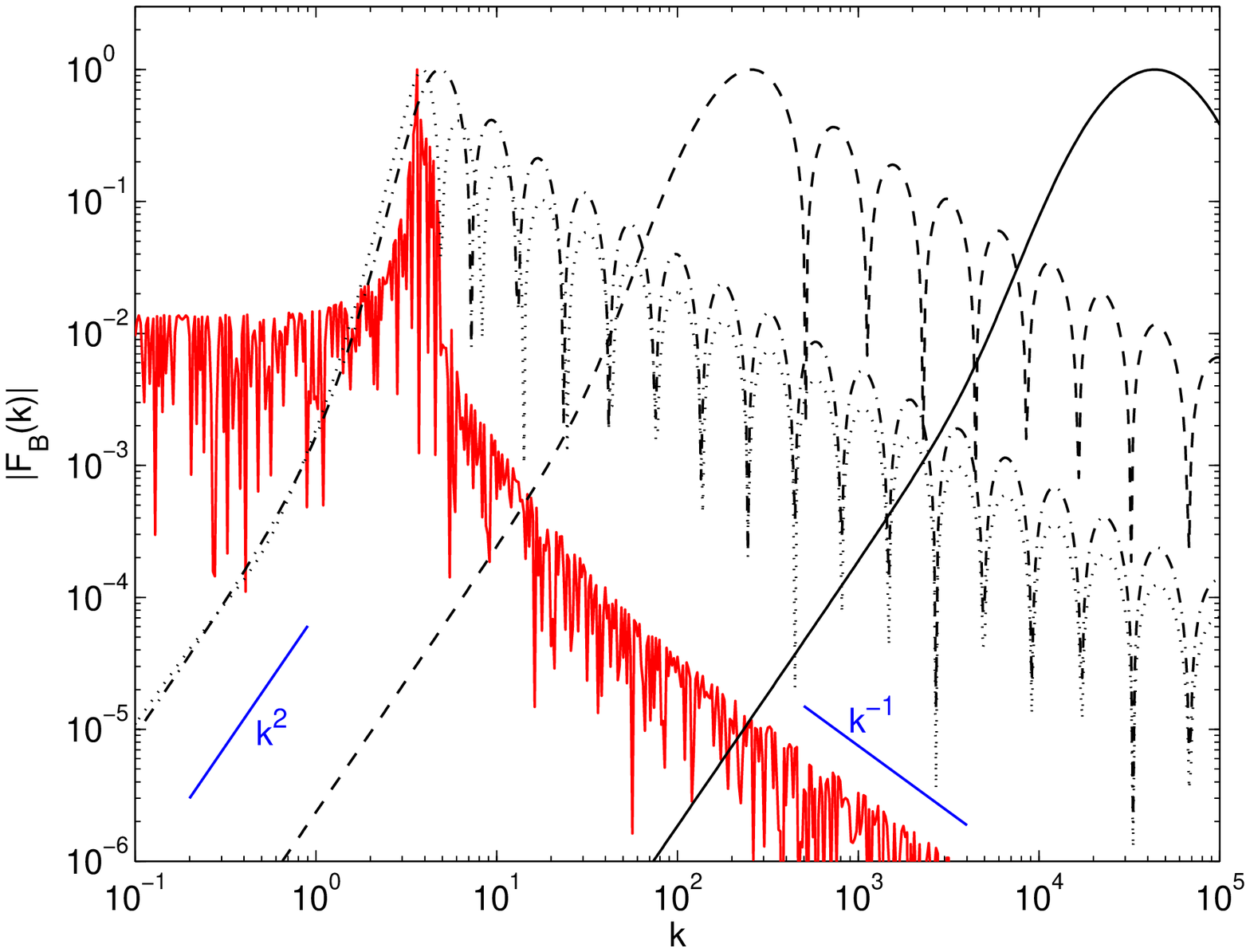}
\includegraphics{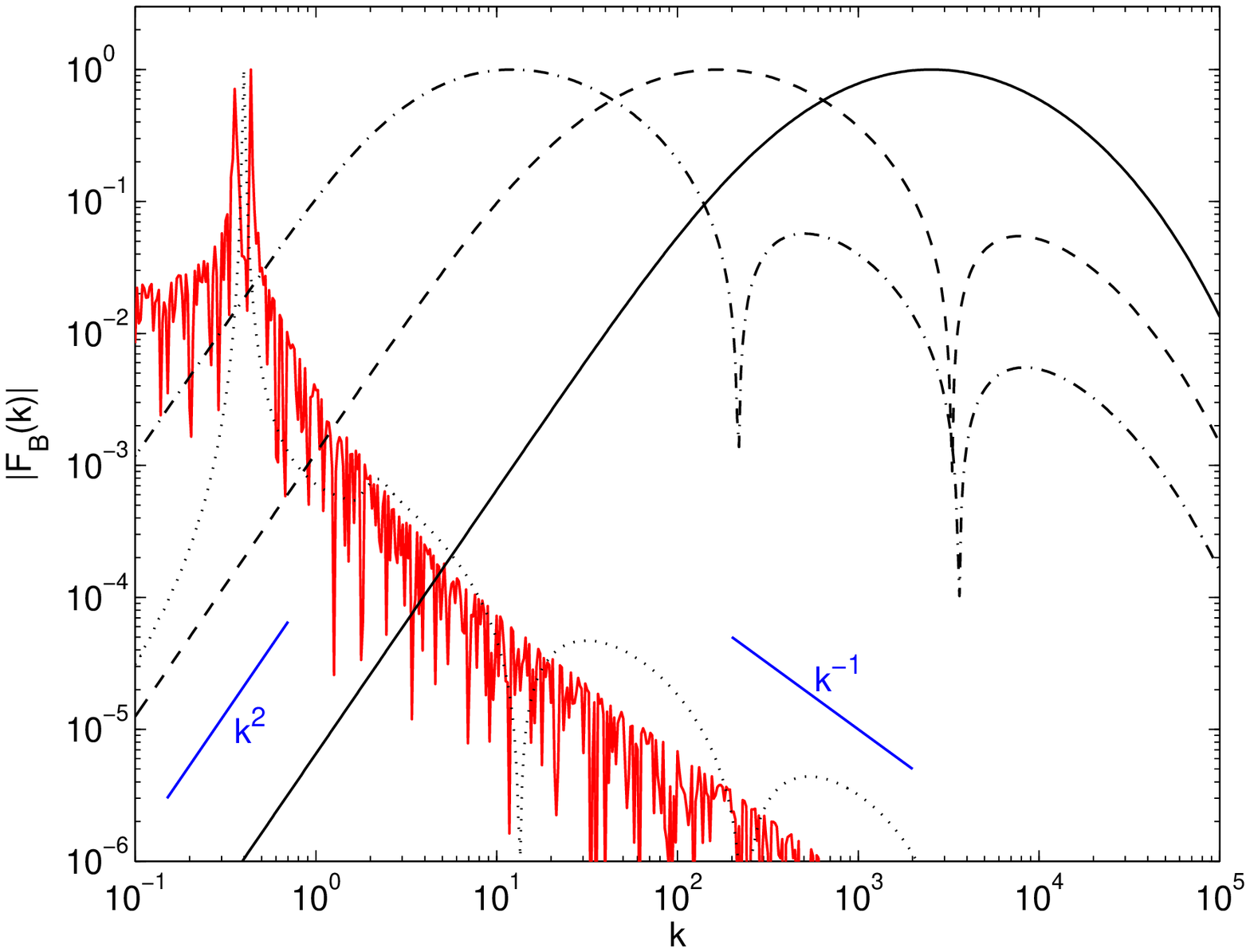}
\caption{Absolute values of magnetic energy spectral eigenfunctions for four selected bound
eigenmodes (shown by the dotted, dash-dotted, dashed, and smooth solid lines),
and for the fastest growing unbound eigenmode (shown by the red jagged spiky solid
lines). The left-upper, right-upper, left-lower, and right-lower plots are for the cases
${\rm Pm}\sim 150$ \& $h=1$, ${\rm Pm}\sim 150$ \& $h=0.1$, ${\rm Pm}\sim 6.7 \times 10^{-4}$ \& $h=1$,
and ${\rm Pm}\sim 6.7 \times 10^{-4}$ \& $h=0.1$, respectively.
\label{FIGURE_2}
}
\end{figure}

\begin{figure}[t]
%\epsscale{1.0}
%\plottwo{f3_UpperLeft_3e3_1.0.eps}{f3_UpperRight_3e3_0.1.eps}
%\plottwo{f3_LowerLeft_3e7_1.0.eps}{f3_LowerRight_3e7_0.1.eps}
\vspace{11.8truecm}
\includegraphics{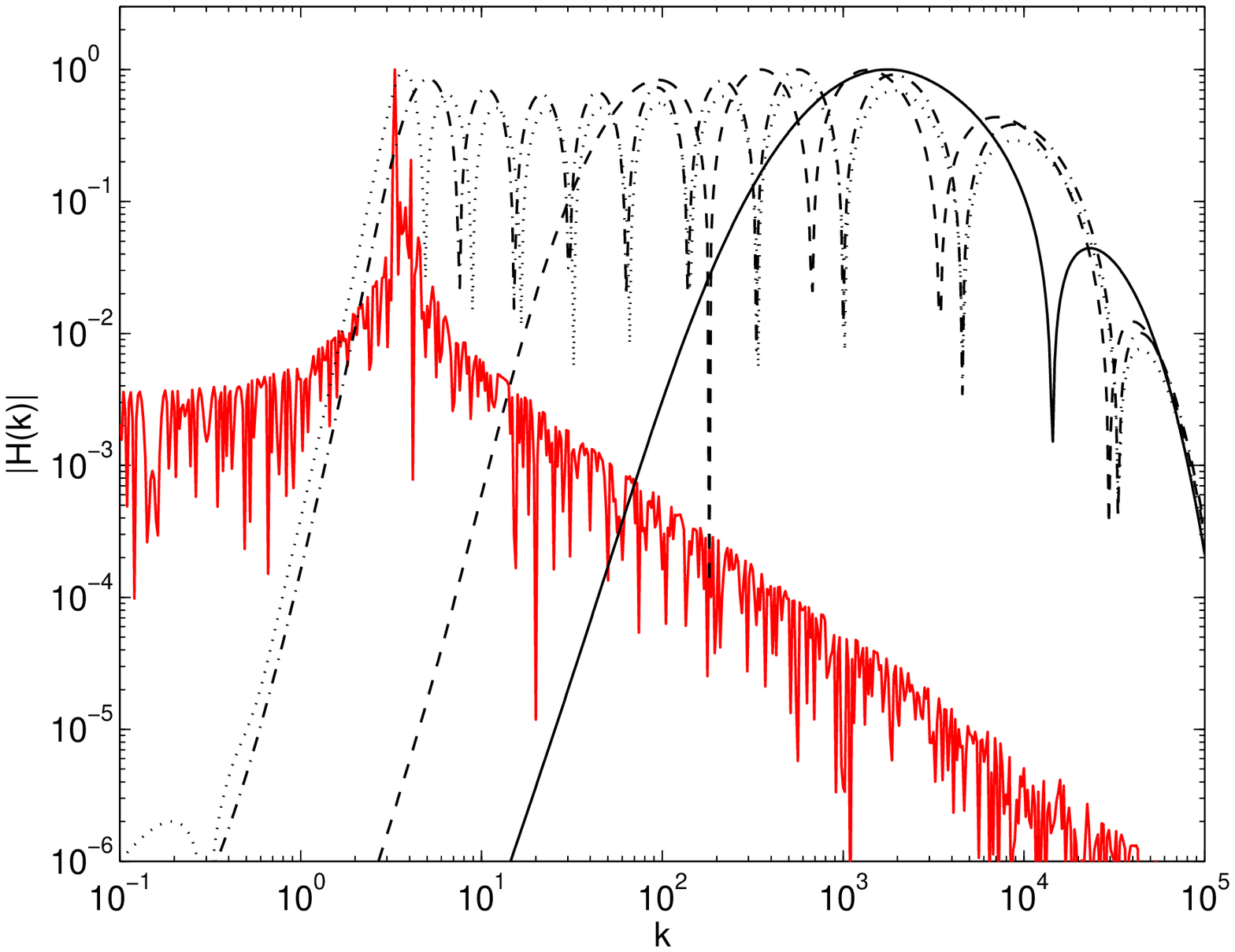}
\includegraphics{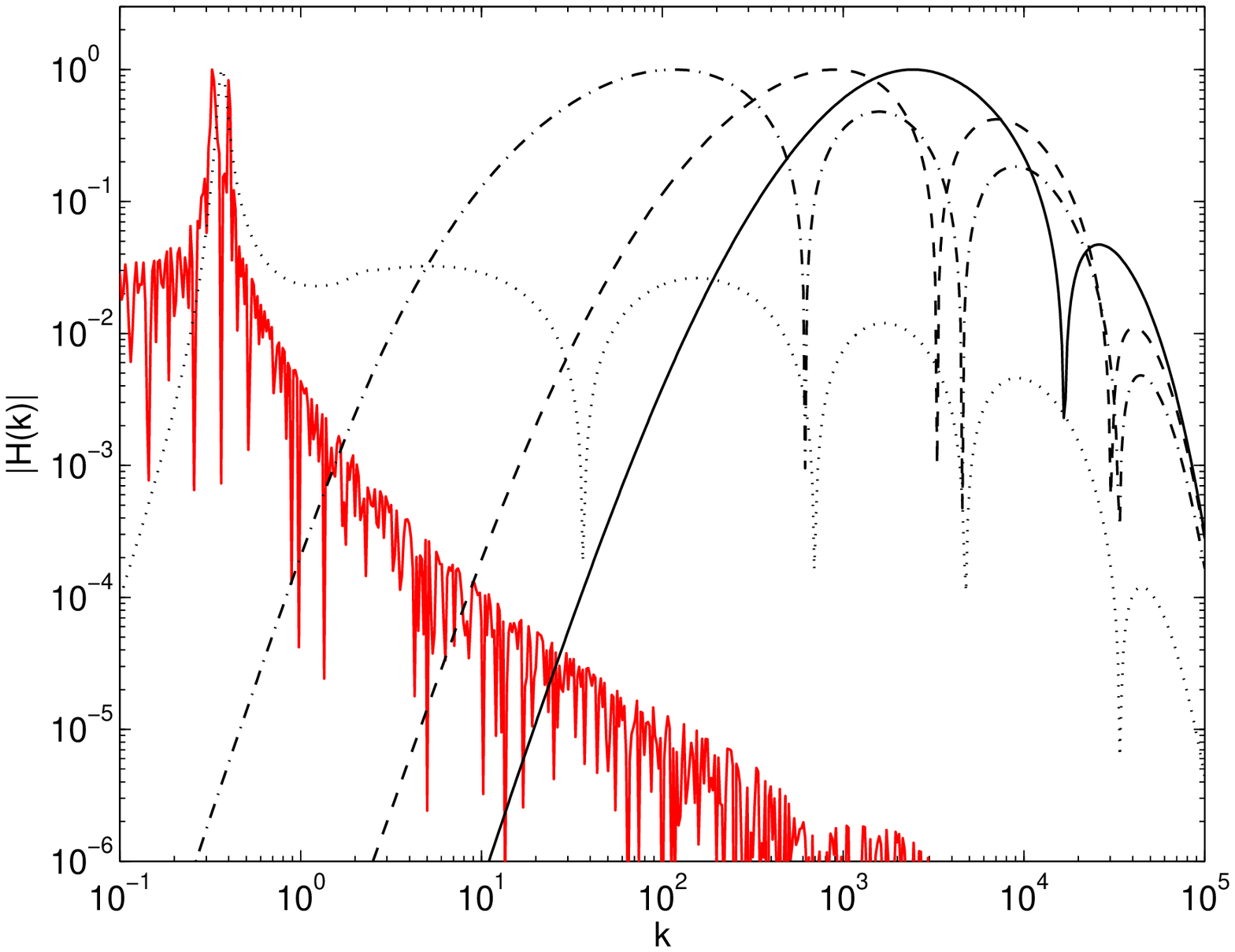}
\includegraphics{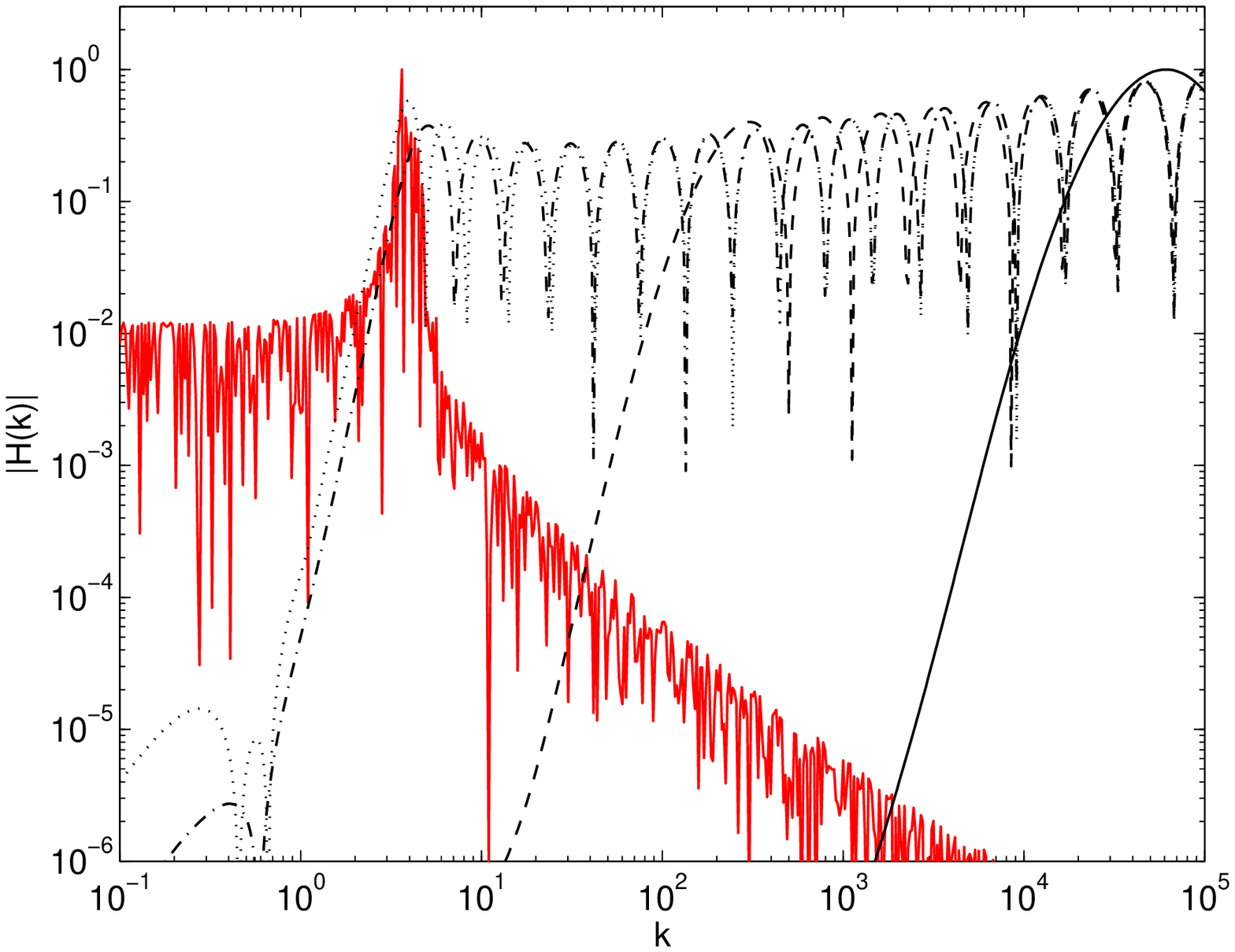}
\includegraphics{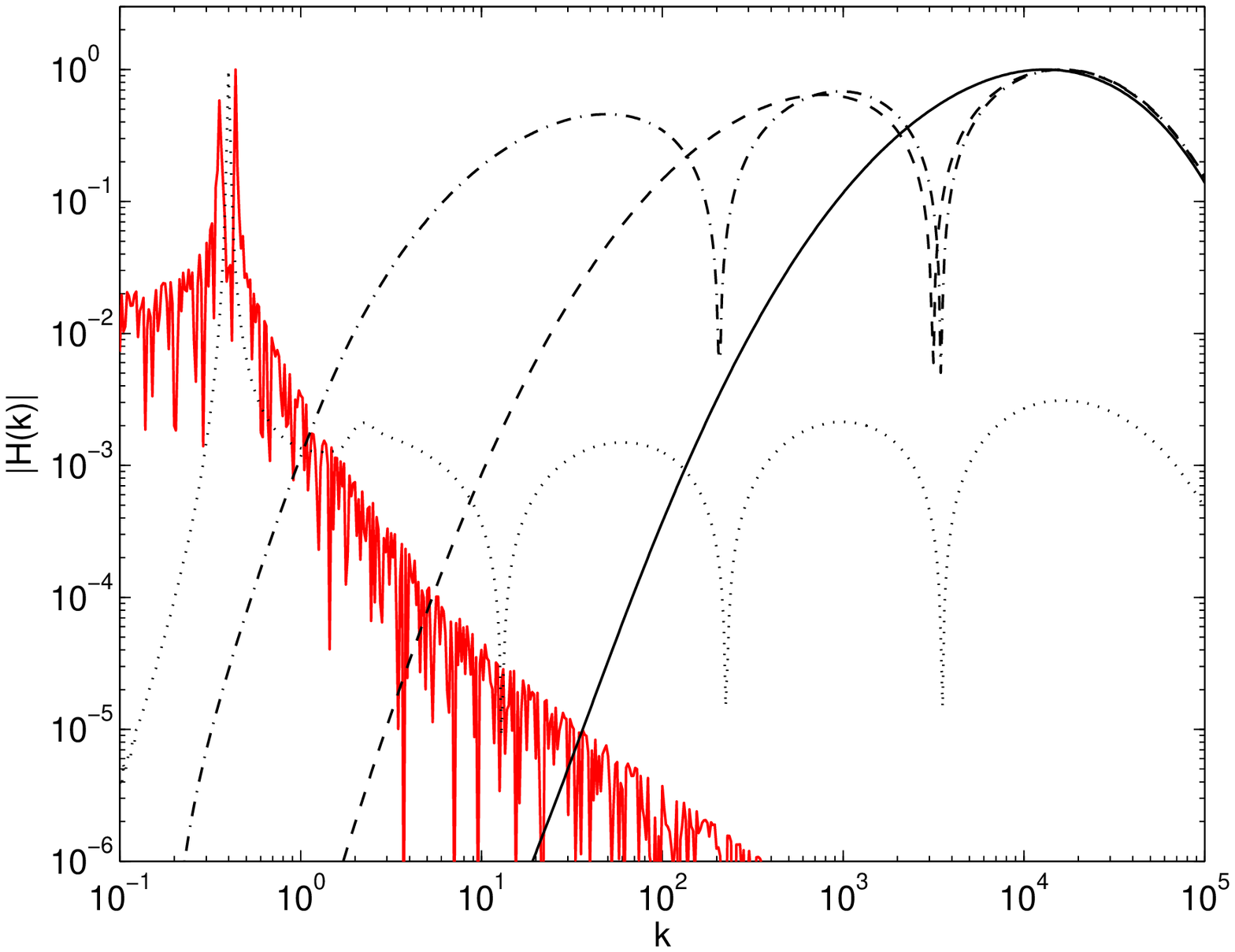}
\caption{Same as Figure~\ref{FIGURE_2} except the absolute values of electric current
helicity spectral eigenfunctions are plotted here.
\label{FIGURE_3}
}
\end{figure}

We choose magnetic diffusivity to be $\eta=10^{-6}$, corresponding to
magnetic Reynolds number ${\rm Rm}\sim 10^6$. By changing
the value of the turbulence cutoff wavenumber $k_{\rm max}$ in Equation~(\ref{POWER_V}),
we vary the Reynolds number ${\rm Re}\approx 1/\nu\approx(k_{\rm max}/4)^{4/3}$ and the 
magnetic Prandtl number ${\rm Pm}={\rm Rm}/{\rm Re}\approx\nu/\eta$. We study two cases
for the Prandtl number: a case when it is large and a case when it is small.
In the first case, ${\rm Pm}\sim 150\gg 1$, which is achieved by choosing $k_{\rm max}=3000$
(${\rm Re}\sim 6800$). In the second case, ${\rm Pm}\sim 6.7 \times 10^{-4}\ll 1$, which corresponds
to our choice $k_{\rm max}=3\times10^7$ (${\rm Re}\sim 1.5\times10^9$).
These two cases of large and small Prandtl numbers are considered in combination
with two cases for the kinetic helicity: first, a case when $h=1$ in Equation~(\ref{POWER_V})
and the velocity field is maximally helical and second, a case when $h=0.1$ and
the kinetic helicity is small. Thus, in total we consider four cases for our choice of
the Prandtl number ${\rm Pm}$ and the kinetic helicity parameter $h$.
The resulting growth rates $\lambda_n$ of the bound (localized) eigenmodes and $\lambda_0$
of the fastest growing unbound (nonlocalized) eigenmode are shown on the logarithmic-scale
plots in Figure~\ref{FIGURE_1}. The growth rates are measured in the units of large-scale
eddy turnover rate~$\sim v_0/l_0$ ($v_0/l_0\sim 1$ here). The logarithmic-scale plots of
the absolute values of magnetic energy spectral eigenfunctions are given in
Figure~\ref{FIGURE_2}. The logarithmic-scale plots of the absolute values of electric
current helicity spectral eigenfunctions are given in Figure~\ref{FIGURE_3}.

In the case ${\rm Pm}\sim 150$ and $h=1$ there exist $10$ growing bound eigenmodes of
magnetic field, whose growth rates are shown on the plot~(A) in Figure~\ref{FIGURE_1}.
Among these we select four bound modes
$\lambda_1\simeq35.41$, $\lambda_2\simeq42.74$, $\lambda_7\simeq213.4$, and $\lambda_{10}\simeq731.1$,
and we plot their magnetic energy and current helicity spectral eigenfunctions by the
dotted, dash-dotted, dashed, and smooth solid lines, respectively, on the left-upper plots in
Figures~\ref{FIGURE_2} and~\ref{FIGURE_3}. The spectral eigenfunctions of the fastest
unbound eigenmode $\lambda_0\simeq33.23$ are shown by the red jagged spiky lines on
these left-upper plots.
In the case ${\rm Pm}\sim 150$ and $h=0.1$ there are just four bound magnetic field
eigenmodes, which are shown on the plot~(B) in Figure~\ref{FIGURE_1}. These eigenmodes are
$\lambda_1\simeq0.3336$, $\lambda_2\simeq26.03$, $\lambda_3\simeq190.1$, and $\lambda_4\simeq654.2$,
and their energy and helicity spectral eigenfunctions are shown by the dotted, dash-dotted,
dashed, and smooth solid lines on the right-upper plots in Figures~\ref{FIGURE_2}
and~\ref{FIGURE_3}. The fastest growing unbound eigenmode grows at a rate
$\lambda_0\simeq0.3323$, and its spectral eigenfunctions are shown by the red jagged
spiky lines.
Next, in the case ${\rm Pm}\sim 6.7 \times 10^{-4}$ and $h=1$ there are eighteen bound magnetic
eigenmodes in total, all are shown on the plot~(C) in Figure~\ref{FIGURE_1}. The spectral
eigenfunctions of four selected bound eigenmodes
$\lambda_1\simeq41.73$, $\lambda_2\simeq49.56$, $\lambda_{10}\simeq695.9$, and $\lambda_{18}\simeq17055$
are shown by the dotted, dash-dotted, dashed, and smooth solid lines on the left-lower plots
in Figures~\ref{FIGURE_2} and~\ref{FIGURE_3}. The spectra of the fastest growing unbound
mode $\lambda_0\simeq39.57$ are again shown by the red jagged spiky
lines.
Finally, in the case ${\rm Pm}\sim 6.7 \times 10^{-4}$ and $h=0.1$ there are four bound eigenmodes,
$\lambda_1\simeq0.39581$, $\lambda_2\simeq12.30$, $\lambda_3\simeq103.5$, and $\lambda_4\simeq669.8$,
refer to the plot~(D) in Figure~\ref{FIGURE_1}. These four modes and the fastest growing
unbound mode $\lambda_0\simeq0.39573$ have spectra that are shown on the right-lower plots
in Figures~\ref{FIGURE_2} and~\ref{FIGURE_3} by the dotted, dash-dotted, dashed, smooth solid,
and red jagged spiky lines, respectively.

Based on the results presented in Figures~{\ref{FIGURE_1}--\ref{FIGURE_3}}, we make the
following important observations.

First, when the kinetic helicity increases, the number of bound magnetic eigenmodes increases
significantly. Their growth rates, $\lambda_n$, become strongly
concentrated near the growth rate of the fastest unbound eigenmode, $\lambda_0$. (This last
result follows from a nearly uniform distribution of $\lambda_n$ on the logarithmic-scale
plots~(A) and~(C) in Figure~\ref{FIGURE_1}.)

Second, on all plots in Figure~\ref{FIGURE_1} the growth rate of the first bound eigenmode,
$\lambda_1$, happens to be very close to $\lambda_0$. We found same result for all other
high Reynolds number cases that we investigated (not reported here), with different admissible
values of the helicity parameter $h$. Thus, we propose that for dynamo driven by high Reynolds number 
Kolmogorov-type velocity field there always exists a shallow bound eigenmode $\lambda_1$, such
that $\lambda_1-\lambda_0\ll\lambda_0$. This shallow mode grows faster than any of the unbound
modes because $\lambda_1>\lambda_0$.

Third, the spectra of the shallow bound eigenmode $\lambda_1$, shown by the dotted 
lines in Figures~\ref{FIGURE_2} and~\ref{FIGURE_3}, and the spectra of the fastest
growing unbound eigenmode $\lambda_0$, shown by the red jagged spiky lines, 
are close near the magnetic energy containing large scales. 
(Small-scale structures of these modes are however different:
the shallow eigenmode has a relatively larger small-scales component.)
In practical applications, the shallow mode grows faster than the unbound modes and 
can dominate at large scales; however it cannot be described by the conventional 
$\alpha$-model for large-scale dynamo, since this model does not capture the
bound modes.     

Fourth, consider the spectra of eigenmodes $\lambda_0$ and $\lambda_1$
(the dotted lines and the red jagged spiky solid lines in
Figures~\ref{FIGURE_2} 
and~\ref{FIGURE_3}). When the value of the kinetic helicity drops by a
factor of $10$ (from $h=1$ to $h=0.1$), the location of the peaks of these spectra
shifts to larger scales by the same factor. Thus, the characteristic scales of
eigenmodes $\lambda_0$ and $\lambda_1$ are both approximately equal to $\sim l_0/h$,
so that both these modes peak at large scale when kinetic helicity is small.~\footnote{
At large correlation scales $x\to \infty$, the eigenfunction of the fastest growing
unbound eigenmode, $\lambda_0=g^2(0)/[\kappa_L(0)+2\eta]\sim h^2v_0/l_0$, asymptotically
becomes a mixture of cosine and sine standing waves with wavenumber
$k_0=\sqrt{\lambda_0}/\sqrt{\kappa_L(0)+2\eta}\sim h/l_0$, while
the bound eigenfunctions decline as ${}\propto \exp(-\mu_n x)$, where 
$\mu_n=\sqrt{\lambda_n-\lambda_0}/\sqrt{\kappa_L(0)+2\eta}$ \citep{boldyrev05,malyshkin}.
} 
This result is consistent with general predictions of the $\alpha$-model~\citep{skr}.

\section{Conclusion}

We have presented the full numerical characterization of the kinematic 
Kazantsev-Kraichnan dynamo model in the case of the Kolmogorov scaling of 
the velocity field. Our main conclusion is that the structure and the
characteristic correlation scales of the magnetic field amplified by helical kinematic
dynamo action are determined by both bound (localized) and unbound (nonlocalized)
growing eigenmodes. In particular, the large-scale component of the
field is defined by the unbound eigenmodes and by the shallow bound eigenmodes.
Because these shallow bound eigenmodes have growth rates higher than those of all
unbound eigenmodes, at any given scale the former may rapidly become dominant over
the latter. In practical applications, this means that the shallow bound modes, rather
than the unbound modes, are likely to become essential in the large-scale magnetic
field configurations in astrophysical systems. In this case the conventional 
$\alpha$-dynamo model~\citep{skr,moffatt78,kulsrud1} gives an inadequate description of
the large-scale magnetic field even at the kinematic stage of dynamo action.

The $\alpha$-model becomes inapplicable in this case because it uses a
critical assumption that small-scale fluctuations 
of the velocity and magnetic fields are much weaker and concentrated at the scales much smaller
than the scales of the growing large-scale field. Under this assumption the $\alpha$-model is
obtained by averaging the induction Equation~(\ref{induction}) over these small-scale fluctuations
to obtain a linear and homogeneous differential equation for the large-scale mean magnetic 
field.\footnote{The $\alpha$-model equation for the mean field $\overline{\bf B}$ is
$\partial_t \overline{\bf B}=\alpha \nabla \times \overline{\bf B}+\beta \nabla^2 \overline{\bf B}$.
This is a linear and homogeneous differential equation with constant coefficients
$\alpha\sim \overline{{\bf v}\cdot ({\nabla \times {\bf v}})} \tau_0$ and $\beta \sim v_0 l_0$,
where $v_0$ is a characteristic velocity, $l_0$ is a characteristic scale,
and $\tau_0\sim l_0/v_0$ is a characteristic decorrelation time of fluid fluctuations.
}
Thus, the $\alpha$-model misses all growing bound magnetic eigenmodes, including the
essential shallow bound eigenmodes that determine the eventual large-scale configuration of
the magnetic field.

We thank Fausto Cattaneo for many useful and stimulating discussions.
This work was supported by the NSF Center for Magnetic Self-Organization in
Laboratory and Astrophysical Plasmas at the Universities of Chicago and
Wisconsin-Madison. S.B.~is supported by the U.S. Department of Energy under
the grant no.~DE-FG02-07ER54932.

%------------------------------------------------------------------------------------------

%------------------------------------------------------------------------------------------

\clearpage

\end{document}